\documentclass[conference]{IEEEtran}
\usepackage{multirow}
\usepackage[ruled, linesnumbered]{algorithm2e}
\usepackage{amsmath}
\usepackage{multirow}
\IEEEoverridecommandlockouts
\usepackage{cite}
\usepackage{amsmath,amssymb,amsfonts}
\usepackage{algorithmic}
\usepackage{graphicx}
\usepackage{amsmath}
\usepackage{textcomp}
\usepackage{xcolor}
\def\BibTeX{{\rm B\kern-.05em{\sc i\kern-.025em b}\kern-.08em
    T\kern-.1667em\lower.7ex\hbox{E}\kern-.125emX}}
\begin{document}

\title{Securing Voice Biometrics: One-Shot Learning Approach for Audio Deepfake Detection}

\author{\IEEEauthorblockN{ Awais Khan$^1$}
\IEEEauthorblockA{\textit{$^1$Department of Computer Science and Engineering,} \\
Oakland University\\
Rochester, Michigan, USA \\
awaiskhan@oakland.edu}
\and
\IEEEauthorblockN{Khalid Mahmood Malik~\IEEEmembership{(Senior Member,~IEEE)}$^{1,2}$} 
\IEEEauthorblockA{\textit{$^1$College of Innovation and Technology,}}
University of Michigan-Flint\\
\IEEEauthorblockA{\textit{$^2$Department of Computer Science and Engineering,}}
Oakland University \\
drmalik@umich.edu, mahmood@oakland.edu}

\maketitle

\begin{abstract}
The Automatic Speaker Verification (ASV) system is vulnerable to fraudulent activities using audio deepfakes, also known as logical-access voice spoofing attacks. These deepfakes pose a concerning threat to voice biometrics due to recent advancements in generative AI and speech synthesis technologies. While several deep learning models for speech synthesis detection have been developed, most of them show poor generalizability, especially when the attacks have different statistical distributions from the ones seen. Therefore, this paper presents Quick-SpoofNet, an approach for detecting both seen and unseen synthetic attacks in the ASV system using one-shot learning and metric learning techniques. By using the effective spectral feature set, the proposed method extracts compact and representative temporal embeddings from the voice samples and utilizes metric learning and triplet loss to assess the similarity index and distinguish different embeddings. The system effectively clusters similar speech embeddings, classifying bona fide speeches as the target class and identifying other clusters as spoofing attacks. The proposed system is evaluated using the ASVspoof 2019 logical access (LA) dataset and tested against unseen deepfake attacks from the ASVspoof 2021 dataset. Additionally, its generalization ability towards unseen bona fide speech is assessed using speech data from the VSDC dataset.
\end{abstract}


\section{Introduction}
\IEEEPARstart{B}{iometrics}, particularly voice-based, holds immense potential for addressing the complex problem of identifying individuals across a variety of devices, e.g., smartphones, voice assistants (Amazon Alexa, Apple Siri, Google Home), etc. As technological advancements continue to transform our everyday transactions and interactions, ensuring reliable authentication mechanisms, such as automatic speaker verification (ASV) systems, becomes increasingly important. However, the security of ASV is vulnerable to recent deepfakes (logical access spoofing), which pose significant risks to the trustworthiness of the systems \cite{b18}.

\begin{figure}[t]
\centerline{\includegraphics[width=\linewidth]{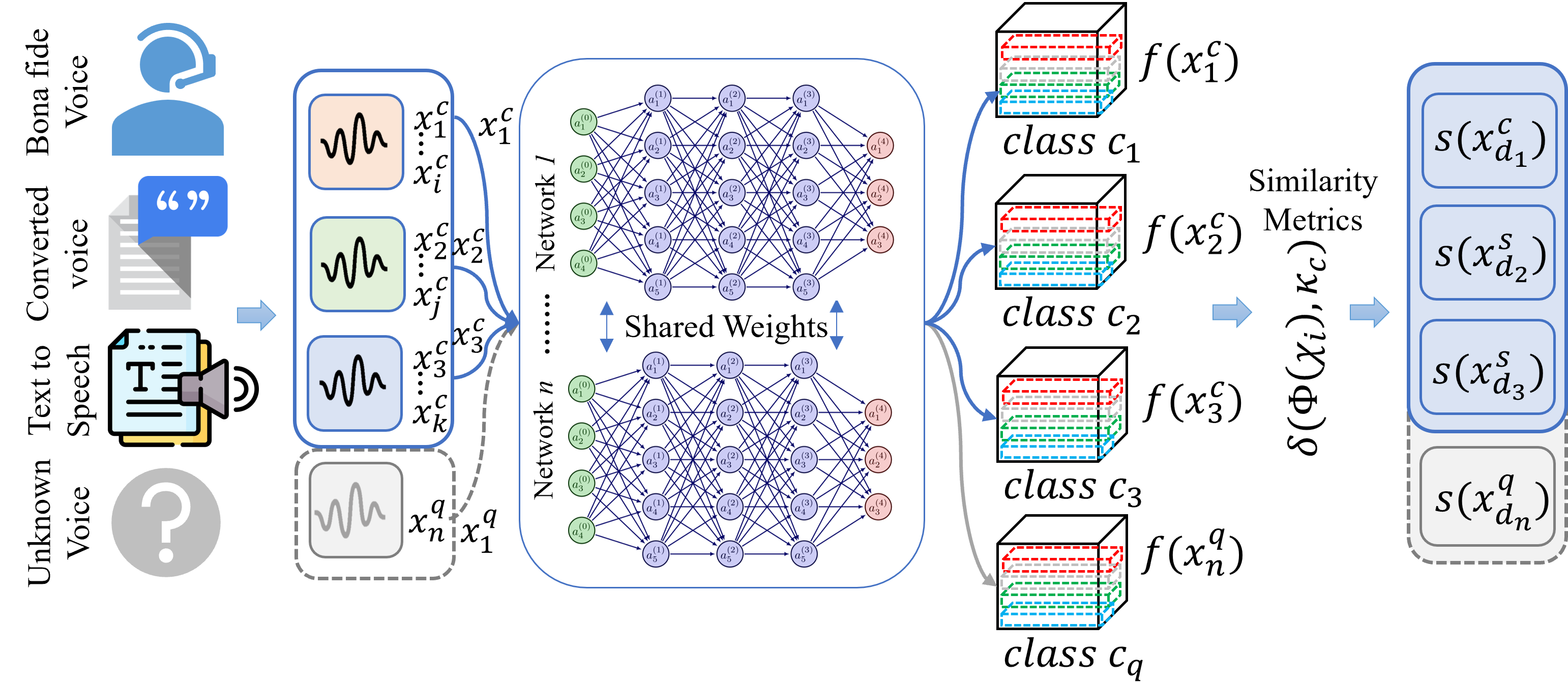}}
\caption{Multiple colors represent various classes. The handcrafted features extracted from audio samples of these classes are used to generate support set samples, denoted as $x^c_1$, $x^c_2$,\dots$x^c_n$. These features are subsequently fed through deep neural networks, resulting in audio embeddings represented as $f(x^c_1)$, $f(x^c_2)$, and $f(x^c_3)$. In parallel, a speech sample from an unknown class (indicated by dotted lines) that requires recognition undergoes a similar processing pipeline, generating query embeddings $f(x^c_q)$, which are then compared with each support class embeddings. Euclidean similarity metrics are employed to compute similarity scores, including $s(x^c_{d_1})$, $s(x^c_{d_2})$, $s(x^c_{d_3})$,\dots$s(x^q_{d_i})$. These similarity scores reflect the likelihood that the query audio belongs to a specific support set class. The final decision is based on selecting the highest similarity score among all comparisons.}
\label{fig1}
\end{figure}

The emergence of generative AI has raised concerns about the potential misuse of Audio Deep-fakes (AD). For instance, Microsoft's latest AI technology can now simulate anyone's voice with just a few seconds of audio \cite{bb2}. Scammers have utilized deep-fake audio to deceive individuals and organizations, resulting in financial losses \cite{b1,b2,b3,bb4}. Examples include a Canadian couple falling victim to a fraudster \cite{b1}, impersonating a CEO, leading to a theft of \$35 million \cite{b2}, and a UK company losing \$243,000 \cite{bb4}. As per the Federal Trade Commission \cite{b3}, Americans alone lost \$2.6 billion last year in impostor frauds, and audio-deepfakes are expected to further enhance such crimes. Additionally, audio deepfakes could erode public trust, manipulate discourse, embarrass individuals, disrupt court proceedings, and lead to the Liar's Dividend phenomenon—where genuine evidence is denied as a deepfake. They may also sway elections, recruit sources through fabricated social media accounts, incite violence, or radicalize populations. These incidents and threats highlight the urgency of developing generalized and trustworthy solutions for deepfake detection.

Considerable efforts have been made to protect voice biometrics against AD attacks \cite{b23,b24,b25,b26,b27,b28,b29,b30}. However, the emergence of new voice cloning algorithms poses significant threats to the security of ASVs. Recent studies highlight the issue of generalization in spoofing detection systems \cite{b5}. Various acoustic features, including LFCC, CQCC, and MFCC, were analyzed on the ASVspoof 2019 dataset, but they show poor generalizability for samples generated using unknown spoofing techniques \cite{b18,b34}. Moreover, existing countermeasures struggle to detect unseen attacks due to the assumption of similar data distributions between training and test utterances, which is incorrect for unseen attacks. Unseen attacks have distinct spectro-temporal artefacts and signature discrepancies that the model is unfamiliar with, leading to misclassification. The distribution mismatch discussed in \cite{b14} is another cause of failure for unseen attacks. In particular, mismatch distribution arises when the number of spoofing samples, generated by various spoofing algorithms, significantly outweighs the number of genuine speech samples. This class imbalance can introduce bias into the system towards the major class of the dataset distribution

The proposed Quick-SpoofNet solution incorporates one-shot and metric learning techniques to address unknown signature discrepancies and mismatched distributions \cite{b14}. One-shot learning compares speeches using a similarity matrix to mitigate the mismatch distribution problem, while metric learning treats spoofing attacks as data points that can appear anywhere in the embedding space, far from the bona fide speech samples. Further, we selected the mel-frequency spectrogram, spectral energy contrast, and spectral shape envelope to facilitate the extraction of effective spectral features from speech samples. The Mel-spectrogram aids in detecting subtle spoofing artefacts; the spectral energy contrast highlights energy distribution variations; and the spectral envelope helps in identifying inconsistencies and abnormalities from the speech sample. More specifically, in the one-shot setting, we utilize LSTM layers in shared networks, which were chosen for their ability to effectively capture temporal dependencies and long-term patterns in the audio data. This enables the model to detect instinctual artefacts and traits indicative of voice spoofing, such as unnatural speech patterns or inconsistent vocal characteristics over time. The dense layer architecture after LSTM layers further refines the vocal embeddings for improved discrimination. In this way, Quick-SpoofNet reduces the need for retraining when encountering unseen attacks, thereby reducing data storage, costs, and system complexity. Our primary focus is on enhancing spoofing detection accuracy with limited data and no prior knowledge of spoofing utterances. The key contributions of this paper are:
\begin{itemize}
\item  We introduce Quick-SpoofNet, a one-shot and metric-based solution for audio deepfake detection. To the best of our knowledge, this is the first-ever evaluation of one-shot and metric learning techniques specifically in the field of voice anti-spoofing.
\item We present a robust feature set that includes the Mel-spectrum, spectral-envelop, and spectral-contrasts, which significantly improve audio deepfake detection. 
\item  By utilizing the presented feature set along with Quick-SpoofNet architecture, we demonstrate generalization capabilities in detecting new and previously unseen deepfake attacks. This enables proactive detection of emerging attacks that did not exist during the development of our solution.
\end{itemize}
The term 'generalization' in our approach signifies its adaptability to detect diverse voice spoofing attacks, accommodating differences in users, unseen attacking techniques, and algorithmic artifacts, as illustrated in Fig. 1. 
\begin{figure*}[htbp]
\centerline{\includegraphics[width=18.5cm]{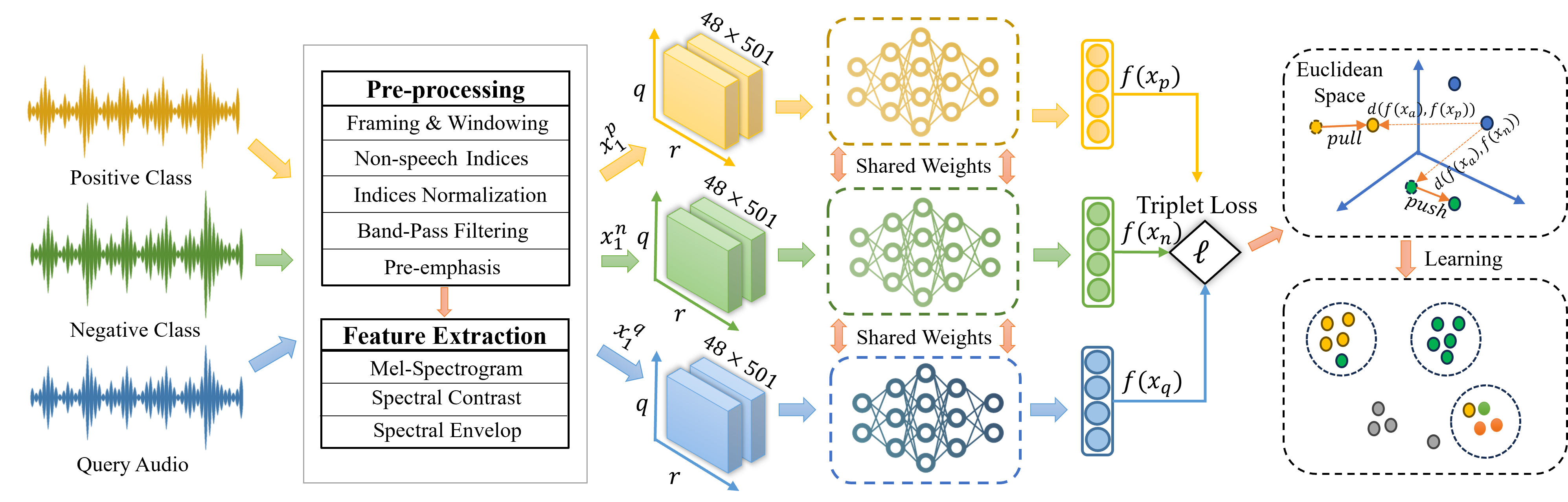}}
\caption{ The proposed system's architectural  diagram illustrates the flow of audio samples represented by multiple colors, each corresponding to a triplet of positive, negative, and query audio. The speech samples undergo initial preprocessing, followed by the extraction of handcrafted features denoted as $x^p_1$, $x^n_2$, and $x^q_3$. These features are then inputted into the shared LSTM-Siamese networks, resulting in vocal embeddings $f(x^c_1)$, $f(x^c_2)$, and $f(x^c_3)$. To classify the speech samples, the triplet loss and Euclidean distance are employed.}
\label{fig2}
\end{figure*}

\section{Literature Review}
This section presents an overview of recent research on identifying synthetically generated voices and unseen attacks. The existing literature can be broadly categorised into two groups: classical machine learning (ML) based countermeasures and deep learning (DL) based solutions.

Classical ML models have been widely used in audio deepfake (AD) detection, achieving high success rates and outperforming other methods \cite{b30,b24,b28,b29}. Despite their high accuracy, existing CNN models rely heavily on preprocessed audio inputs, leading to the development of deep learning (DL) methods that automate feature extraction and reduce preprocessing requirements. In fake multimedia detection, the research community developed various audio deepfake classification models \cite{b23,b26,b25,b27,b13}. However, these systems faced a limitation in generalization due to their strong dependency on the training speech samples, making them ineffective in detecting unseen attacks.

In addressing the generalization limitations of synthetic audio detection, a one-class strategy based on speaker verification was used in synthetic audio detection \cite{b16}. In another study, to handle unknown classes, the solution was tested against open-set and closed scenarios and showed superior performance compared to the baseline systems \cite{b17}. Alegre et al. \cite{b8} showed the effectiveness of a one-class-based support vector machine (OC-SVM) that trained entirely on bona fide speech samples. Zhang et al. introduced a loss function, known as "one-class softmax" \cite{b14}, to distinguish genuine voices from counterfeit audio samples. These studies motivate us to avoid relying on the inherent characteristics of real and spoof data and instead present a strategy that can detect seen and unseen attacks by analysing the similarity index.

Furthermore, one-shot learning solutions based on Siamese networks have gained attention in computer vision applications like person recognition and object classification \cite{b35}. However, its potential in speech analysis and voice biometrics remains relatively unexplored. Some limited work has been done using one-shot learning for speaker identification \cite{b19,b20,b21,b22}. Velez et al. \cite{b19} present the Siamese convolutional neural network (SCNN) for speaker identification using one-shot classification. Elof et al. \cite{b20} investigated Siamese representations in multimodal one-shot learning, combining spoken and visual numbers for diverse audio classification. Other researchers, including Manocha et al. \cite{b21} and Zhang et al. \cite{b22}, introduced content-based approaches using Siamese CNNs, resulting in improved performance compared to existing frameworks. Despite the limited work on one-shot learning for speaker identification, further exploration and advancements are needed to leverage this approach in voice analysis and deep-fakes detection.

\section{Methodology}
\subsection{Proposed Method}
As shown in Fig. \ref{fig2}, the proposed method uses a Siamese network architecture with shared weights and employs one-shot learning and metric-learning techniques to extract robust voice temporal embeddings. These embeddings are then utilized in the Euclidean latent space to differentiate between bona fide and spoofed speech. The training process involves dividing the dataset into three subsets: a training subset $(\tau)$, a support subset $( \delta )$, and a query subset $(\varphi)$. While the classes (encompassing both spoof and bona fide) remain the same across all subsets, the speakers/users in the training $(\tau)$, support $(\delta)$, and query datasets $(\varphi)$ are distinct.

The $\tau$ subset is used to train models in a metric-learning setup, while $\delta$ and $\varphi$ contain speech samples that are different from the training subset $\tau$. The training subset $\tau$ is comprised of arbitrary pairs of speech samples and their target labels from the $X_i$ classes. Let the speech sample from training subset $\tau$  be represented as $\Delta_\tau$. Thus, the training subset includes $\tau = {(x^t_i, y^t_i)}^{\Delta_\tau}_{i=1}$ speech samples.

In one-shot-based learning architectures, the support set $\delta$ comprises $X_j$ distinct classes, each employing a single pair of speech samples and their corresponding target labels $\delta = {(x^\delta_j, y^\delta_j)}^{K_\delta}$ where ${j=1}$. Let the speech sample from support subset $\delta$ be represented as $K_\delta$.The primary objective of this approach is to classify unseen speech samples and specify the target label $ y_{\varphi}$ for any $x^\varphi_i$. In other words, the goal is to assign labels to query voices as bona fide or spoof by comparing their similarity with the support set. Once the dataset is split into three subsets, the next step involves pre-processing and feature extraction, resulting in a robust feature set with dimensions $(q\times r)$, represented as $(X^p_i,X^n_i,X^\varphi_i)$.

The pre-processing and feature extraction methods, detailed in our previous work \cite{b13}, enhance shared network efficiency through five steps. Initially, speech samples are segmented into frames, followed by non-silence speech retrieval, amplitude normalization, band-pass filtering, and pre-emphasis filtering. After obtaining the highly representative pre-process speech frames, feature extraction is performed. The feature extraction process includes log Mel-spectrogram, spectral envelope, and spectral contrast speech features, which have demonstrated effectiveness in known spoofing detection \cite{b13}. For further details on the pre-processing and feature extraction processes, we refer readers to our previous work in \cite{b13}.

In the third step, the feature set $(X^p_i,X^n_i,X^\varphi_i)$ obtained from the last phase is further passed to the shared LSTM-based Siamese networks as input speech features. Next, by employing triplet loss, three speech samples are simultaneously processed by the shared network to extract robust temporal vocal embeddings represented as $f(x_p)$, $f(x_n)$, and $f(x_\varphi)$. 
\begin{equation}
s(x^c_{d_i})_p= \phi(f(x^t_{a_i}), f(x^t_{p_i}))
\end{equation}
\begin{equation}
s(x^c_{d_i})_n= \phi(f(x^t_{a_i}), f(x^t_{n_i}))
\end{equation}
where the $\phi$ operation is used to measure the similarity between the embeddings of anchor $f(x^t_{a_i})$, negative $f(x^t_{n_i})$, and positive $f(x^t_{p_i})$ speech samples. Specifically, it quantifies the closeness between the embeddings of the anchor sample and the positive or negative samples. The function $f(x^t_i)$, representing the embeddings of the speech samples, is continuously updated with respect to the triplet loss to enhance the robustness of vocal traits extracted from the feature set of each speech sample. The equation used for triplet loss is as follows:
\begin{equation}
    L_{triplet} = max(0,\parallel s(x^c_{d_i})_p \parallel ^{2}_{2} - \parallel s(x^c_{d_i})_n \parallel ^{2}_{2}+ \alpha)
\end{equation}
where $\left|\right|^2_2$ denotes the Euclidean norm. $\alpha$ is a margin hyper-parameter that determines the desired separation between the anchor-positive pair and the anchor-negative pair in the embedding space. This loss function encourages the embeddings of anchor-positive pairs to be closer to each other than the embeddings of anchor-negative pairs by at least a margin $\alpha$.

In the final step, the Euclidean distance similarity metric creates a latent space to differentiate between speech samples based on the dissimilarity of their embeddings. Positive speech samples are drawn closer to the query speech sample when their distance is minimal, while negative speech samples move away when their distance is maximal. This approach effectively classifies query samples by evaluating the distances between their embeddings, addressing both unseen and seen spoofing attacks.

\subsection{Testing}
During the testing phase, the query subset comprises multiple speech samples from various classes that have not been seen during training, while the support set consists of single samples from the same classes as the query set. For the testing of the proposed solution, the features are extracted from query samples, and then the obtained query feature set is input to the shared LSTM-based Siamese network to obtain the query embedding. Lastly, we compare the obtained query embeddings with the embeddings of each speech sample in the support subset. Later, the decision will be made based on the highest probabilistic similarity to the query speech sample. This process is repeated for all remaining query samples, following the same technique. The overall testing flow is visualized in Fig. \ref{fig1}.

Specifically, the proposed network predicted embeddings for evaluation query speech samples. We calculated the distance between the query sample and the embeddings of the speech samples in the support set to classify the utterance as either genuine or spoofed. Particularly if the distance exceeded the threshold $t$, a label of 1 was assigned; otherwise, 0 was assigned. By comparing predicted and ground-truth labels, model accuracy and Equal Error Rate (EER) were computed.


\section{Experiments and Results}
In this section, we present the experimental setup, evaluation, and results conducted to evaluate the performance of our proposed approach in detecting audio deepfakes. 

\subsection{Setup}
\subsubsection{Dataset and Evaluation Metrics}
The proposed method was evaluated using the ASVspoof2019 dataset \cite{b32}, specifically the LA subpart, which contains 22,800 spoofed and 2,580 genuine speech samples from various spoofing algorithms. The evaluation split of ASVspoof2019 \cite{b32} serves as an in-domain evaluation dataset with unseen attacks, but it shares configuration characteristics with the training dataset, limiting model generalization assessment. Therefore, to assess generalization, subsets of the ASVspoof2021-DF \cite{b33} and VSDC-0PR \cite{b31} datasets were used, representing unseen real-world scenarios. The VSDC-0PR \cite{b31} subset had 1008 genuine speech samples, while ASVspoof2021-DF \cite{b33} included 25,000 randomly selected deepfake speech samples. Performance was measured using EER, accuracy, F1-score, precision, and recall as assessment metrics.

\subsubsection{Implementation details and Hyper-parameters}
The proposed solution was developed on a system with four NVIDIA Tesla V100 $16$GB GPUs, $192$GB of RAM, and $48$ core CPUs operating $@ 2.10$GHz. The LSTM-based Siamese network has an input shape of $(48, 501)$ comprised of a spectrum feature set. It consists of two LSTM layers with $64$ nodes, followed by dense layers of sizes $512$, $256$, and $128$. Batch normalization and a dropout of $0.2$ are applied. The network uses a batch size of $64$ speech samples and the ReLU activation function for most layers, while the last dense layer uses the sigmoid activation function. The triplet loss function employs a margin alpha value of $0.2$. The Adam optimizer is used with an initial learning rate of $0.001$ and a schedule for learning rate decay every $5000$ steps. A threshold value of $t = 1 \times 10^{-3}$ is used for computing predicted labels. 

\subsection{Results}

\subsubsection{Performance Evaluation of the Proposed Method against Synthetic voice Attacks in ASVspoof2019}
In this experiment, we tested the performance of Quick-SpoofNet against the ASVspoof2019-LA dataset. The presented system obtained an overall EER of 0.50\% and an accuracy of 98.5\%. We test the performance of Quick-SpoofNet separately against VC (voice cloning) and voice conversion algorithms. In testing voice cloning, the efficacy of the proposed solution was evaluated against SOTA VC algorithms in the dataset ranges from A01 to A06, and the results are shown in Table \ref{tab:clonningresults}. The model attained EER values of 0.07\%, 0.15\%, 0.06\%, 0.17\%, 0.02\%, and 0.09\% for A01 to A06 algorithms, respectively, showing optimum performance against each VC algorithm.

Next, we tested the performance of the system against unknown spoofing traits, and the results are shown in Table \ref{tab:unseenresults}. The proposed method performed effectively against several forms of spoofing attacks, delivering a reduced EER and greater accuracy. Despite the fact that the error rate for A17-A19 algorithms was slightly higher than for the rest of the spoofing techniques due to distinct speech characteristics, the proposed method performed overall better than the rest of the spoofing algorithms. This illustrates the system's ability to generalise and identify unknown and unseen speech spoofing attempts efficiently.

\begin{table}[htbp]
  \caption{Performance evaluation of the proposed solution against voice clonning algorithms of ASVspoof2019 dataset.}
  \label{tab:clonningresults}
   \begin{center}
  \begin{tabular}{|c|c|c|c|c|c|} \hline
   \textbf{Algorithm} & \textbf{EER\%} &\textbf{ Acc\%} & \textbf{Precision\%} & \textbf{Recall\%} &\textbf{F1\%}  \\
\hline
   A01 & 0.07 & 98.7 & 93.76 & 95.60 & 94.34 \\ \hline
   A02 & 0.15 & 93.1 & 92.86 & 92.85 & 92.78 \\ \hline
   A03 & 0.06 & 99.3 & 93.63 & 94.25 & 93.66 \\ \hline
   A04 & 0.17 & 94.6 & 90.90 & 89.56 & 91.25 \\ \hline
   A05 & 0.02 & 96.3 & 94.88 & 94.23 & 95.36 \\ \hline
   A06 & 0.09 & 93.0 & 91.32 & 91.33 & 90.32 \\ \hline
\end{tabular}
\end{center}
\end{table}

\begin{table}[htbp]
  \caption{Performance evaluation of the proposed system against unseen voice spoofing algorithms.}
  \label{tab:unseenresults}
   \begin{center}
  \begin{tabular}{|c|c|c|c|c|c|c|}
\hline
   \textbf{Algorithm} & \textbf{EER\%} & \textbf{Acc\%} & \textbf{Precision\%} & \textbf{Recall\%} & \textbf{F1\%}  \\
 \hline
   A07 & 0.25 & 98.50 & 97.29 & 93.20 & 95.50 \\ \hline
   A08 & 0.31 & 97.30 & 98.36 & 92.55 & 95.42 \\ \hline
   A09 & 7.30 & 92.35 & 82.69 & 80.25 & 79.66 \\ \hline
   A10 & 0.30 & 97.50 & 97.90 & 92.40 & 93.25 \\ \hline
   A11 & 0.20 & 98.30 & 95.88 & 91.23 & 93.36 \\ \hline
   A12 & 0.21 & 98.90 & 93.32 & 92.33 & 91.17 \\ \hline
   A13 & 0.50 & 96.95 & 98.79 & 94.62 & 94.78 \\ \hline
   A14 & 0.71 & 96.13 & 97.72 & 95.43 & 95.18 \\ \hline
   A15 & 2.40 & 93.34 & 96.17 & 94.23 & 95.66 \\ \hline
   A16 & 1.45 & 95.62 & 94.70 & 92.74 & 92.25 \\ \hline
   A17 & 20.53 & 89.18 & 65.88 & 60.23 & 63.36 \\ \hline
   A18 & 30.57 & 70.15 & 75.90 & 73.56 & 72.56 \\ \hline
   A19 & 25.30 & 85.93 & 52.88 & 40.23 & 45.36 \\ \hline
\end{tabular}
\end{center}
\end{table}
\begin{figure*}[t]
\centerline{\includegraphics[width=\linewidth]{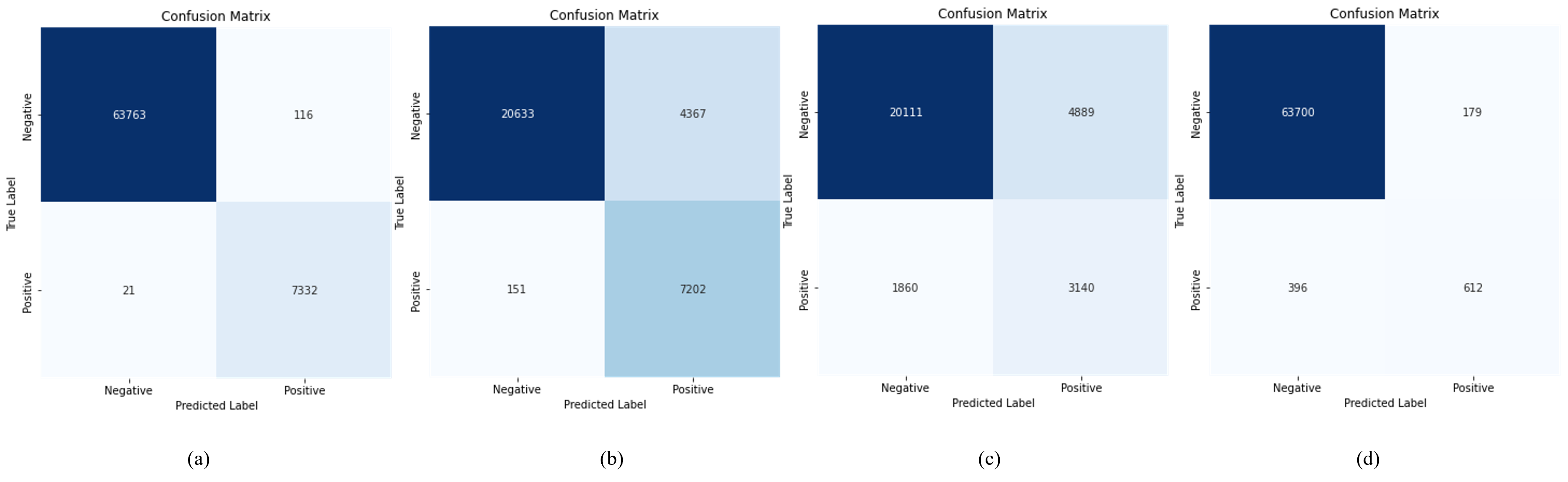}}
\caption{ The confusion metrics of the proposed solution were evaluated for both seen and unseen spoofing attacks. Figure (a) represents the performance of the solution on the complete ASVspoof2019 dataset. Figure (b) illustrates the performance against unseen spoofing attacks, where bona fide speeches are sourced from ASVspoof2019 and attacking samples are sourced from ASVspoof2021. Figure (c) showcases the performance against both unseen real and unseen spoofing attacks, with real samples obtained from VSDC and spoofing samples from ASVspoof2021. Lastly, Figure (d) demonstrates the performance against unseen real speech samples, where real speech samples are acquired from VSDC and spoofing samples from ASVspoof2019.}
\label{fig3}
\end{figure*}

\subsubsection{Performance Evaluation against Unseen Speech Samples and Cross Corpus}
The performance of the Quick-SpoofNet solution was evaluated on unseen speech, and the results are shown in Fig. \ref{fig3} and Table \ref{tab:crossdataset}. The proposed solution demonstrated higher accuracy when tested on speakers present in the training datasets. However, its performance varied when tested on unseen bonafide and spoofing attacks. The solution effectively detected spoofing attacks from the ASVspoof2021-DF datasets, with a slight decrease in performance. It also performed well with unregistered, bona fide speakers. Overall, the proposed solution effectively detects bona fide and state-of-the-art spoofing speeches when the speaker is a registered entity in the training dataset. Nevertheless, the model's performance declined when tested on cross-dataset unseen real and spoofing speech samples together.

\begin{table}[htbp]
  \caption{Performance evaluation of the proposed solution on unseen bonafide and spoofing datasets and cross corpora.}
  \label{tab:crossdataset}
   \begin{center}
  \begin{tabular}{|c|c|c|c|c|c|} \hline
   \multicolumn{2}{c}{\textbf{Dataset}}  & \multicolumn{3}{c}{\textbf{Performance Metrics}} & \\ \hline
   \textbf{Bona fide} & \textbf{Spoofing} &\textbf{ Acc\%} & \textbf{Pre\%} & \textbf{Rec\%} &\textbf{F1\%}  \\
\hline
   ASVspoof19 & ASVspoof19 & 98.9 & 98.29 & 97.20 & 98.12 \\ \hline
   ASVspoof19 & ASVspoof21 & 86.41 & 82.36 & 99.55 & 90.42 \\ \hline
   VSDC & ASVspoof21 & 77.63 & 80.25 & 91.66 &  85.80\\ \hline
   VSDC & ASVspoof19 & 98.50 & 98.90 & 95.56 & 97.25 \\ \hline
\end{tabular}
\end{center}
\end{table}

\subsubsection{Comparative performance analysis with the SOTA baselines and comparative methods}
In this experiment, we conducted a performance comparison between the proposed solution and nine recent works that employed feature fusion for detecting deepfake attacks. The results, shown in Table \ref{tab:performance_fusioncomparison}, indicate that the presented solution outperformed other comparative feature fusion-based methods. Notably, our solution exhibits superior performance compared to our previous work, despite utilizing similar feature sets. This demonstrates the effectiveness of robust vocal embeddings extracted from the LSTM-based Siamese network and the post-distance-based similarity metrics for speech classification.

\begin{table}[htbp]
  \caption{Performance comparison of Quick-SpoofNet with comparative feature-fusion methods.}
  \label{tab:performance_fusioncomparison}
  \begin{tabular}{|c|c|c|c|}
     \hline
   \textbf{Study} &  \multicolumn{2}{c}{\textbf{Fusion System}} &  \\  \hline
      --  & \textbf{Classifier} & \textbf{Frontend Features} & \textbf{EER} \\
       \hline
    2019 \cite{b23} & ResNet & MFCC, Spec, CQCC & 6.02  \\ \hline
    2019 \cite{b24} & LCNN & CQT,LFCC,FFT, CMVN+LFCC &  1.84 \\ \hline
    2020 \cite{b25} & DenseNet & spec,LFCC  & 1.98  \\ \hline
    2021 \cite{b26} & Capsule & LFCC, STFT-gram  & 1.07 \\ \hline
    2021 \cite{b27}& SE-ResNet50 & Spec,LFCC,CQT  & 1.89 \\ \hline
    2021 \cite{b28}& SENet & dual-band of FFT &  1.56  \\ \hline
    2022 \cite{b29}& LCNN+GMM & CQCC,LFCC, RLFCC  & 2.57  \\ \hline
    2022 \cite{b30}& scDenseNet & Spec,LFCC,ARS  & 1.01 \\\hline
    2022 \cite{b30}& scDenseNet & SpecL,LFCC,ARS & 0.98 \\ \hline
    2022 \cite{b13}& SpotNet & Mel-spect, Spec-Env and Cons & 0.95 \\ \hline
    \textbf{Proposed }& One shot learning & Mel-spect, Spec-Env and Cons & 0.50  \\ \hline
  
\end{tabular}
\end{table}
\section*{Limitation and Future Work}
The results in Table \ref{tab:crossdataset} show a decline in accuracy when the presented solution is tested against unseen bona fide speech from different datasets, like VSDC \cite{b31}. This discrepancy occurs due to the strong dependence and comparison of embeddings from registered speakers bona fide samples, which leads to misclassification when confronted with unknown speakers bona fide speech with distinct artefacts. In contrast, the method performs well against unseen spoof samples, as their embeddings differ greatly from those of registered bona fide speakers. This highlights the effectiveness of the proposed method in detecting spoofing against registered ASV speakers. Additionally, the solution is specifically tested against deepfake audio, but we plan to expand its application to other types of spoofing attacks in the future, such as adversarial and physical access attacks. Furthermore, we aim to address generalization limitations towards real speakers and we will explore Transformers as a temporal feature extraction method within a unified learning framework.

\section*{Conclusion}
This study investigates one-shot learning and metric learning for audio deep-fake detection, including unseen attacks. Quick-SpoofNet shows potential for identifying seen and unseen deep-fake audio from different datasets. It achieves an EER of 0.50\% on ASVspoof2019 for seen spoofing attacks and 86.41\% accuracy for unseen attack detection. Moreover, it reaches 98.50\% accuracy for bona fide speech from unseen speakers. Although there was a slight decline in performance during cross-corpus testing for unseen genuine speeches, the proposed solution demonstrated its effectiveness and generalization capability against previously unseen spoofing attacks targeting ASV speakers.

\section*{Acknowledgment}
This presented work is supported by the National Science Foundation (NSF) under award numbers 2329858 \& 2231619 and Michigan Transnationals Research and Commercialization (MTRAC), Advanced Computing Technologies (ACT) award number 292883.


\end{document}